\ifx\onecol\undefined
\documentclass[journal,twocolumn]{IEEEtran}
\else 
\documentclass[12pt,journal,onecolumn]{IEEEtran}
\fi
\usepackage{booktabs}
\usepackage{amsmath,amsfonts,amssymb}
\usepackage{algorithmic}
\usepackage{algorithm}
\usepackage{array}
\usepackage{color}
\usepackage{diagbox}
\usepackage{textcomp}
\usepackage{stfloats}
\usepackage{url}
\usepackage{verbatim}
\usepackage{graphicx}
\graphicspath{{Figures/}}
\usepackage{cite}
\usepackage{amsthm}
\usepackage{gensymb}
\usepackage{bm}
\usepackage{setspace}
\usepackage{tabularx}
\usepackage{hyperref}
\hypersetup{hidelinks, 
colorlinks=true,
allcolors=black,
pdfstartview=Fit,
breaklinks=true}
\usepackage{subfigure}     

\begin{document}


\title{Site-Specific Cell Shaping: An Agentic Framework}


\author{Zhuo Xu, \textit{Graduate Student Member, IEEE}, Zhaolin Wang, \textit{Member, IEEE}, \\ and Yuanwei Liu, \textit{Fellow, IEEE} \vspace{-2em}
\thanks{

The authors are with the Department of Electrical and Computer Engineering, The University of Hong Kong, Hong Kong (e-mails: \href{mailto:ecezhuoxu@connect.hku.hk}{ecezhuoxu@connect.hku.hk}; \href{mailto:zhaolin.wang@hku.hk}{zhaolin.wang@hku.hk}; \href{mailto:yuanwei@hku.hk}{yuanwei@hku.hk}). }
}



\maketitle

\begin{abstract}

An agentic framework is proposed for cell shaping in future sixth-generation (6G) wireless network planning. Cell shaping controls the coverage and interferences of base stations (BSs) through antenna parameters configuration, providing an important means of improving coverage. In contrast to conventional approaches that reply on costly performance evaluations, the proposed framework achieves intelligent and efficient antenna configuration through a perception-reasoning-reflection loop.
Specifically, a \textit{Configuration Agent} combines scene information, evaluation history, and reflection feedback to generate antenna configuration proposals. 
Then, a \textit{ray tracing(RT)-based evaluator} assesses these proposals and returns a structured performance summary.
Next, a \textit{Reflection Agent} interprets these results to diagnose site-specific performance deficiencies, recommend feasible configuration adjustments, and accumulate conditional experience.  
Through this iterative process, the framework accumulates site-specific experience and progressively refines antenna configurations without human-in-the-loop intervention.
Simulations in real-world urban scenarios demonstrate that the proposed framework achieves lower outage probabilities than the considered benchmark methods, highlighting the effectiveness of agentic reasoning and reflection in translating site-specific performance feedback and accumulated experience into adaptive antenna configuration decisions.

\end{abstract}

\begin{IEEEkeywords}
	Agents for communications, cell shaping, site-specific, large language models, 6G.
\end{IEEEkeywords}

\section{Introduction}

Sixth-generation (6G) mobile communications are envisioned to support immersive services and pervasive intelligence through closer integration of the physical and digital worlds~\cite{saad2019vision,itu2023imt2030,lu2024tutorial,cui2022near}. Realizing this vision requires 6G wireless networks to provide ubiquitous and reliable connectivity while meeting increasing demands for data rates and spectrum efficiency. In cellular networks, achieving these objectives depends critically on the spatial distribution of desired signals and inter-cell interference~\cite{wang2024tutorial,zhang2023mixed,xu2026how}. \emph{Cell shaping} refers to the adjustment of cell coverage footprints through the configuration of base station (BS) antenna radiation patterns, including the control of antenna downtilt angles as well as horizontal and vertical beamwidths. These configurations jointly influence the signal strength within each cell and the interference imposed on neighboring cells, thereby determining the tradeoff between coverage and capacity~\cite{ning2026broadcasting}. By coordinating antenna configurations across cells, cell shaping can better align coverage with user demand, mitigate coverage holes and excessive inter-cell interference, and improve cell-edge service quality. It thus provides an important means of enhancing network capacity and coverage through more effective utilization of existing wireless infrastructure.

The effectiveness of cell shaping is inherently \emph{site-specific}, since the coverage footprint generated by a given antenna configuration depends on its interaction with the local propagation environment~\cite{wang2026generative,zhao2026site,zhou2026beam}. In practical deployments, building geometry, terrain, and electromagnetic material properties govern blockage, reflection, diffraction, and scattering, producing spatially irregular coverage and interference patterns. Consequently, identical antenna settings can yield substantially different network performance across deployment sites. For example, when a user cluster is obstructed by buildings, increasing antenna gain along the direct geometric direction may offer limited improvement, whereas directing energy toward a viable reflected path can enhance the received signal strength~\cite{hoydis2023sionna}. Moreover, these propagation mechanisms can also strengthen interference at users in neighboring cells, making the impact of antenna adjustments dependent on the surrounding network geometry. These dependencies motivate site-specific cell shaping, in which antenna configurations are tailored to the actual three-dimensional environment, BS locations, and user distribution, with their effects on desired signals and inter-cell interference evaluated jointly~\cite{tekgul2024joint}.

Translating these site-specific requirements into practical antenna configurations remains challenging. Conventional radio-frequency (RF) planning relies on propagation simulations and engineering expertise, while subsequent updates often involve field measurements and trial-and-error tuning~\cite{luo2008dynamic}. Repeated measurements and manual revisions incur substantial time and operational costs, limiting the configurations that can be explored. Moreover, inter-cell interference couples antenna settings, so isolated adjustments may degrade neighboring cells. Such workflows scale poorly and provide no guarantee of near-optimal performance. These limitations motivate automated, sample-efficient optimization that incorporates site-specific propagation information and identifies effective antenna configurations with fewer evaluations and less human intervention.

\subsection{Prior Works}
Existing studies on automated cell shaping commonly address coverage and capacity optimization (CCO) through antenna parameter tuning. These methods can be broadly grouped into three categories. 

The first comprises \textit{analytical and heuristic approaches}, which optimize antenna configurations using explicit network models or prescribed search procedures. Specifically, the authors of \cite{eckhardt2011vertical} proposed a heuristic gradient-ascent method for autonomous antenna tilt adjustment, improving both average and cell-edge spectral efficiency in LTE networks. The authors of~\cite{partov2015utility} formulated antenna tilt optimization as utility maximization and developed a distributed primal--dual algorithm. In~\cite{phan2017coverage}, the authors employed modified particle swarm optimization for load-aware antenna tilt adjustment, increasing the number of served users and network throughput. Besides, a quantization-theoretic framework for jointly optimizing antenna tilts and transmit powers was proposed in \cite{karimi2024optimizing}.

The second comprises \textit{reinforcement learning (RL)-based approaches}, which learn antenna control policies through interactions with the network environment. Specifically, the authors of \cite{razavi2010fuzzy} combined fuzzy logic with reinforcement learning to achieve distributed antenna downtilt optimization in LTE networks without synchronization among cells. The authors of \cite{balevi2019online} integrated offline multi-agent mean-field learning with online single-agent reinforcement learning to jointly adjust antenna tilts and horizontal and vertical beamwidths in heterogeneous networks. In \cite{jin2022graph}, a graph attention Q-learning-based scheme was proposed to incorporate neighboring-cell information into tilt control, improving performance and generalization across network deployments. Besides, \cite{skocaj2022cellular} combined minimization of drive tests (MDT) measurements, electromagnetic simulations, and network performance indicators to train a deep Q-network for antenna tilt optimization.

The third comprises \textit{surrogate-assisted optimization} approaches, including Bayesian optimization (BO) and surrogate-guided evolutionary search. The authors of \cite{dreifuerst2021optimizing} applied multi-objective BO to jointly tune transmit powers and antenna downtilts, with substantially fewer evaluations. In \cite{tekgul2024joint}, the authors combined Gaussian process surrogate models with differential evolution for site-specific antenna tilt and beamwidth tuning, jointly optimizing uplink and downlink coverage and capacity. More recently, the authors of \cite{benzaghta2025cellular} investigated high-dimensional BO for joint antenna tilt and beamwidth optimization while examining transfer learning across deployment scenarios.

However, these methods still have several limitations. Analytical methods typically require simplifying assumptions for tractability, whereas heuristic searches can incur substantial evaluation costs without guaranteeing global optimality. RL-based methods may require extensive environment interactions and converge slowly, resulting in costly training when each interaction involves a high-fidelity network simulation. Surrogate-assisted methods depend on predictive accuracy, which can deteriorate when deployment conditions differ from those represented in the training data. In particular, conventional BO with exact Gaussian process inference incurs a computational cost of $\mathcal{O}(N^3)$ for $N$ observations, while increasing the number of antenna parameters further complicates surrogate learning and acquisition optimization, limiting scalability without additional structural assumptions or algorithmic adaptations.

\subsection{Our Contributions}

Recent advances in large language model (LLM)-based agents have demonstrated substantial potential across information retrieval, programming, and interactive decision-making by integrating reasoning, tool use, and reflection on environmental feedback~\cite{yao2023react,shinn2023reflexion}. In wireless communications, these capabilities enable agents to interpret network observations and task objectives, invoke domain-specific simulators and optimization tools, and generate configuration decisions informed by the operating environment~\cite{liang2026adaptation,hu2026llm,jiang2026from}. By incorporating domain knowledge and retaining experience from previous evaluations, agents offer a promising approach to adapting configuration strategies to site-specific conditions while reducing reliance on manual analysis and adjustment~\cite{li2026comagent,li2026agenticwirelesscommunication6g,zhou2026agents}.

In this paper, we propose an agentic framework for cell shaping, which integrates scene information, realizable antenna configurations, and ray-tracing feedback into a perception--reasoning--reflection loop to minimize downlink outage probability. Specifically, our contributions are summarized as follows:
\begin{itemize}
    \item We propose an agentic framework for site-specific cell shaping that jointly configures the electrical downtilts and horizontal and vertical half-power beamwidths (HPBWs) of BS antennas to minimize downlink outage probability. The framework incorporates scene information, evaluated network performance, and accumulated experience into a perception--reasoning--reflection loop. By interpreting coverage deficiencies and interference patterns, the agents formulate configuration hypotheses and revise their decisions using feedback. This approach enables automated antenna adjustment tailored to the propagation environment.

	\item We first construct a pattern-validated antenna configuration library that associates realizable downtilt and beamwidth combinations with explicit array excitation weights. Building on this feasible action space, a Configuration Agent generates proposals through diagnostic adjustment, coordinated reconfiguration, and systematic probing, with deterministic validation and duplicate removal before evaluation. A ray tracing(RT)-based evaluator assesses the proposals and returns outage performance together with SINR, regional coverage, and interference statistics. A Reflection Agent interprets these results to identify site-specific performance deficiencies, recommend feasible adjustments, and extract conditional experience. These components jointly establish a closed-loop architecture connecting configuration generation, numerical evaluation, and feedback-guided refinement.
    
	\item  We develop a closed-loop optimization procedure that coordinates the two agents and the RT evaluator through a shared state containing the best retained configuration, evaluation history, reflection feedback, and site-specific experience. At each iteration, evaluated outcomes guide subsequent proposals, completed adjustment tests update the experience memory, and the best evaluated configuration is preserved. The procedure operates within a prescribed evaluation budget and terminates according to iteration, stagnation, or proposal-availability conditions. Simulations in real-world urban scenarios demonstrate lower outage probabilities than the considered benchmarks, supporting the effectiveness of combining scene information with agentic reasoning and iterative reflection for antenna configuration.

\end{itemize}

\subsection{Organization and Notation}

The rest of this paper is organized as follows. Section \ref{sectionII} introduces the system model and problem formulation. In Section \ref{sectionIII}, the proposed agentic framework for antenna configurations is introduced. Simulation results and conclusions are presented in Section \ref{sectionIV} and Section \ref{sectionV}. Besides, the notation of this paper is summarized in Table \ref{tab:main_notations}.

\begin{table*}[!t]
\caption{Summary of main notations}
\label{tab:main_notations}
\centering
\footnotesize
\renewcommand{\arraystretch}{1.12}
\setlength{\tabcolsep}{4pt}
\begin{tabularx}{\textwidth}{
    @{}
    >{\raggedright\arraybackslash}p{0.16\textwidth}
    >{\raggedright\arraybackslash}X
    >{\raggedright\arraybackslash}p{0.16\textwidth}
    >{\raggedright\arraybackslash}X
    @{}
}
\toprule
\textbf{Symbol} & \textbf{Definition}
& \textbf{Symbol} & \textbf{Definition} \\
\midrule

$B,\mathcal{B};\,U,\mathcal{U}$
& Numbers and sets of BSs and UEs, respectively
& $\mathcal{Q}_t,K_t$
& Proposal batch and its size at iteration $t$ \\

$\mathbf{p}_b,\mathbf{q}_u$
& Positions of BS $b$ and UE $u$
& $\mathcal{D}_{t,k}$
& Structured antenna configuration proposal $k$ at iteration $t$ \\

$M,P_b,W$
& Number of array elements, BS transmit power, and bandwidth
& $\mathbf{x}_{t,k},\mathbf{x}_{t,k}^{\mathrm{ref}}$
& Proposed and reference network antenna assignments \\

$\theta_b,\beta_b^{\mathrm{h}},\beta_b^{\mathrm{v}}$
& Electrical downtilt and horizontal and vertical HPBWs of BS $b$
& $\mathbf{y}_{t,k},J_{t,k}$
& Structured performance summary and outage probability of a proposal \\

$\mathbf{x}_b,\mathbf{x}$
& BS antenna parameter vector and network configuration; $\mathbf{x}$ uses library indices in Section~III
& $\mathcal{Y}_t$
& Evaluated proposal batch with performance summaries \\

$\mathbf{w}_b(\mathbf{x}_b),G_b$
& Array excitation vector and realized antenna power gain of BS $b$
& $\mathcal{D}_t^\star,\mathbf{y}_t^\star$
& Best proposal in the current batch and its performance summary \\

$\mathbf{h}_{ub}[k]$
& Element-level channel vector from BS $b$ to UE $u$
& $\mathcal{C}_t^{\mathrm{ref}}$
& Input context of the Reflection Agent \\

$P_{\mathrm{out}}(\mathbf{x})$
& Network downlink outage probability
& $\mathcal{R}_t$
& Reflection output comprising diagnoses, guidance, and new experience \\

$\mathcal{C}$
& Finite set of admissible antenna configurations
& $\mathcal{I}_t,\mathcal{G}_t$
& Site-specific diagnoses and configuration adjustment guidance \\

$\mathcal{L},L$
& Validated parameter--weight library and number of entries
& $\mathcal{M}_t,\Delta\mathcal{M}_t$
& Accumulated site-specific experience and its increment \\

$\hat{\mathbf{a}}_{\ell},\mathbf{w}_{\ell}$
& Realized antenna parameters and excitation vector of library entry $\ell$
& $\mathcal{H}_t,\mathcal{A}_t$
& Evaluation history and recommendation execution records \\

$\ell_b$
& Library entry assigned to BS $b$
& $\mathcal{B}_t$
& Best retained proposal and its evaluation record \\

$t,T$
& Iteration index and maximum number of iterations
& $\Xi_t$
& Shared optimization state at iteration $t$ \\

\bottomrule
\end{tabularx}
\end{table*}

\section{System Model and Problem Formulation}
\label{sectionII}
In this section, the system model is first introduced. Then, we formulate the cell shaping problem.

\subsection{System Model}
\label{subsec:system_model}
We consider a downlink cellular network comprising $B$ small-cell base stations (BSs) and $U$ single-antenna user equipments (UEs). The BS and UE sets are denoted by $\mathcal{B}=\{1,\ldots,B\}$ and $\mathcal{U}=\{1,\ldots,U\}$, respectively. The positions of BS $b$ and UE $u$ are given by $\mathbf{p}_b$ and $\mathbf{q}_u\in\mathbb{R}^{3\times1}$. Each BS is equipped with a directional uniform planar array (UPA) containing $M$ antenna elements, transmits with a fixed total conducted power $P_b$ over a bandwidth $W$, and employs a single coverage beam and continuously transmits mutually independent signals. It is assumed that the BS locations, mechanical orientations, transmit powers, array geometries, and UE locations remain fixed throughout the optimization. Cell shaping is carried out by adjusting the electrical downtilt and beamwidths of UPAs of each BS.

\begin{figure*}[t!]
	\centering
	\includegraphics[width=0.95\textwidth]{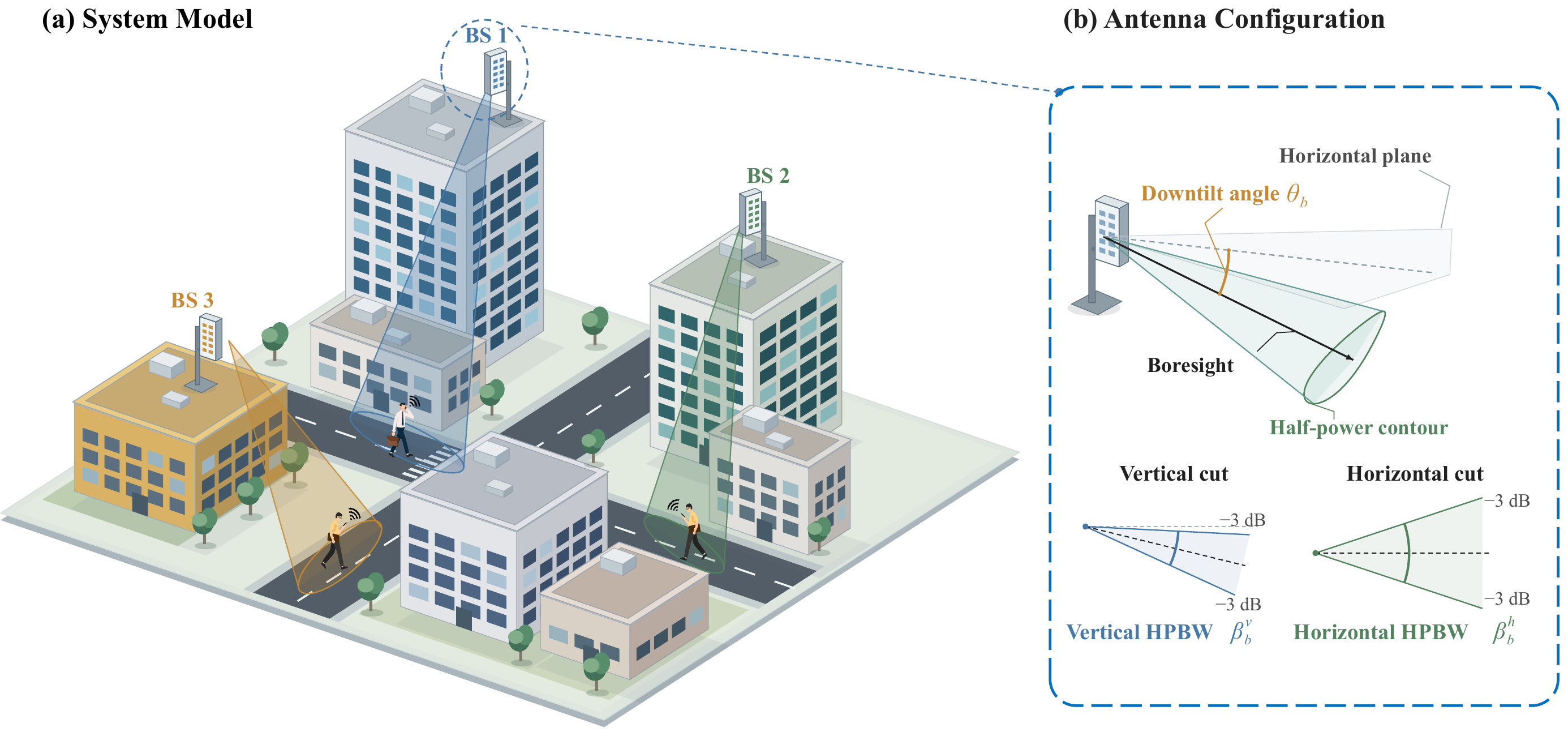}
	\caption{Illustration of the system model and the antenna configuration of BS antennas. (a) system model; (b) antenna configuration}
	\label{Systemmodel}
        \vspace*{-5mm}
\end{figure*}

As shown in Fig.~\ref{Systemmodel}, the antenna configuration of BS $b$ is described by
\begin{equation}
    \mathbf{x}_b =
    \left[\theta_b,\beta_b^{\mathrm{h}},\beta_b^{\mathrm{v}}\right]^{T},
    \qquad
    \mathbf{x} =
    \left[\mathbf{x}_1^{T},\ldots,\mathbf{x}_B^{T}\right]^{T},
    \label{eq:antenna_configuration}
\end{equation}
where $\theta_b$ is the electrical downtilt angle, and $\beta_b^{\mathrm{h}}$ and $\beta_b^{\mathrm{v}}$ are the horizontal and vertical HPBWs, respectively. Let $\alpha$ denote the elevation angle positive above the horizontal plane, and $\varphi$ denotes the azimuth angle relative to the fixed panel orientation. Thus, the antenna gain can be expressed as~\cite{tekgul2024joint}
\begin{equation}
\begin{aligned}
    T_b(\alpha,\varphi)
    &= 10^{-\frac{1}{10}
    \min\left\{
    A_b^{\mathrm{h}}(\varphi)+A_b^{\mathrm{v}}(\alpha),
    A_{\max}
    \right\}},
\end{aligned}
\label{eq:target_pattern}
\end{equation}
where $A_b^{\mathrm{v}}(\alpha)= 12\left(\frac{\alpha+\theta_b}{\beta_b^{\mathrm{v}}}\right)^2$ and $A_b^{\mathrm{h}}(\varphi)= 12\left(\frac{\varphi}{\beta_b^{\mathrm{h}}}\right)^2$. $A_b^{\mathrm{h}}$ and $A_b^{\mathrm{v}}$ are attenuation terms in dB, and $A_{\max}$ is a prescribed attenuation cap. 

The attenuation model in~\eqref{eq:target_pattern} specifies the
desired panel-pattern shape for the requested downtilt and HPBWs.
To realize this target using a finite antenna array, each supported
configuration $\mathbf{x}_b$ is mapped to a deterministic excitation
vector $\mathbf{w}_b(\mathbf{x}_b)\in\mathbb{C}^{M\times1}$ selected from a validated configuration library, which will be detailed in Section \ref{subsectionconfigurationagent}. The weights satisfy $\|\mathbf{w}_b\|_2^2=1$, ensuring that $P_b$ represents the total panel transmit power. Assuming identical, vertically polarized elements and negligible mutual coupling, the realized antenna power gain can be expressed as~\cite{lebret1997antenna}

\begin{equation}
\begin{aligned}
    G_b(\alpha,\varphi;\mathbf{x}_b)
    ={}& G_{\mathrm{e}}(\alpha,\varphi)\\
    &{}\times
    \left|
    \sum_{m=1}^{M}
    w_{b,m}(\mathbf{x}_b)
    e^{j\frac{2\pi}{\lambda}
    \mathbf{r}_m^{T}\mathbf{u}(\alpha,\varphi)}
    \right|^2,
\end{aligned}
\label{eq:realized_antenna_gain}
\end{equation}
where $G_{\mathrm{e}}$ is the fixed element power pattern specified in~\cite[Table~7.3-1]{3gpp2024tr38901}, $\lambda$ is the carrier wavelength, $\mathbf{r}_m$ is the position of element $m$ in the panel coordinate system, and $\mathbf{u}(\alpha,\varphi)$ is the unit vector pointing in the observation direction. Thus, the target pattern guides the synthesis of $\mathbf{w}_b(\mathbf{x}_b)$, while $G_b(\alpha,\varphi;\mathbf{x}_b)$ describes the radiation pattern actually produced by these weights.


Let $\mathbf{h}_{ub}[k]\in\mathbb{C}^{M\times1}$ denote the complex channel from BS $b$ to UE $u$ at frequency sample $k$, which includes the  transmit-element radiation response in (\ref{eq:realized_antenna_gain}). For $K$ equally weighted frequency samples across the operating bandwidth and a flat transmit power spectral density, the received power is approximated by
\begin{equation}
    q_{ub}(\mathbf{x}_b)
    =
    \frac{P_b}{K}
    \sum_{k=1}^{K}
    \left|
    \mathbf{h}_{ub}^{H}[k]\mathbf{w}_b(\mathbf{x}_b)
    \right|^2.
    \label{eq:received_power}
\end{equation}

For each network configuration, UE $u$ associates with the BS providing the largest average received power. The corresponding BS index is given by 
\begin{equation}
    b_u
    =
    \arg\max_{b\in\mathcal{B}}
    q_{ub}(\mathbf{x}_b).
    \label{eq:association}
\end{equation}
Thus, the downlink SINR can be expressed as 
\begin{equation}
    \gamma_u(\mathbf{x})
    =
    \frac{
    q_{u b_u}(\mathbf{x}_{b_u})
    }{
    \displaystyle
    \sum_{b\in\mathcal{B}\setminus\{b_u\}}
    q_{ub}(\mathbf{x}_b)
    +\sigma^2
    },
    \label{eq:downlink_sinr}
\end{equation}
where $\sigma^2$ is the receiver noise power. Antenna reconfiguration changes the desired signal, inter-cell interference, and potentially the serving BS, coupling the configuration decisions across the network.

Cell shaping aims to provide reliable wireless coverage within a prescribed service region. A key metric for evaluating coverage is the outage probability, defined as the probability that a UE sampled from the service region experiences an SINR below a prescribed threshold. Given a prescribed SINR threshold $\gamma_{\mathrm{th}}>0$, the network outage probability can be expressed as
\begin{equation}
    P_{\mathrm{out}}(\mathbf{x})
    =
    \frac{1}{U}
    \sum_{u=1}^{U}
    \mathbf{1}
    \left\{
    \gamma_u(\mathbf{x})<\gamma_{\mathrm{th}}
    \right\},
    \label{eq:outage_probability}
\end{equation}
where $\mathbf{1}\{\cdot\}$ is the indicator function, which equals one when the condition inside the braces is satisfied and zero otherwise.


\subsection{Problem Formulation}
\label{subsec:problem_formulation}

Our objective is to minimize the downlink outage probability by jointly configuring the antenna downtilt and horizontal and vertical HPBWs of all BSs. Let $\mathcal{L}$ denote the finite library of validated antenna configurations introduced in Section~\ref{subsectionconfigurationagent}, where each entry specifies a parameter triple and its associated array vector. With the BS locations, transmit powers, and propagation environment fixed, the cell shaping problem can be formulated as
\begin{subequations}
\label{eq:cell_shaping_problem}
\begin{align}
    \underset{\mathbf{x}}{\operatorname{minimize}}
    \quad &
    P_{\mathrm{out}}(\mathbf{x})
    =
    \frac{1}{U}\sum_{u=1}^{U}
    \mathbf{1}\left\{
    \gamma_u(\mathbf{x})<\gamma_{\mathrm{th}}
    \right\}
    \label{eq:cell_shaping_objective}\\
    \operatorname{subject~to}
    \quad &
    \mathbf{x}_b\in\mathcal{L},
    \qquad b\in\mathcal{B}.
    \label{eq:cell_shaping_constraint}
\end{align}
\end{subequations}
where
$\theta_b\in[\theta_{\min},\theta_{\max}]$,
$\beta_b^{\mathrm{h}}\in[\beta_{\min}^{\mathrm{h}},\beta_{\max}^{\mathrm{h}}]$,
and
$\beta_b^{\mathrm{v}}\in[\beta_{\min}^{\mathrm{v}},\beta_{\max}^{\mathrm{v}}]$ with strictly positive minimum HPBWs. Additionally, it should be noted that these adjustable antenna parameters must satisfy the array power and radiation-pattern requirements, which will be elborated in detail in Section \ref{subsectionconfigurationagent}. The above parameter bounds are used during library construction, whereas Problem~\eqref{eq:cell_shaping_problem} optimizes over the resulting finite validated library.

Problem (\ref{eq:cell_shaping_problem}) is nonconvex and combinatorial. The antenna configurations jointly affect the desired and interfering powers, while configuration-dependent UE association further couples the decisions across BSs. If each BS has $L=|\mathcal{L}|$ admissible configurations, the joint search space contains $L^B$ possibilities, making exhaustive enumeration impractical as the network grows.

In addition, compared with exhaustive enumeration, existing learning-based and surrogate-based methods can reduce the evaluation burden, but they face different challenges. RL-based approaches often require task-specific policy training or retraining, whereas conventional scalar-objective BO-based approaches do not explicitly translate spatial evaluation feedback into testable adjustment guidance and traceable configuration experience. These challenges motivate the proposed agentic framework, which will be discussed in the following section.






\section{Proposed Agentic Framework}
\label{sectionIII}
In this section, we propose an agentic framework for site-specific cell shaping, comprising a \emph{Configuration Agent} and a \emph{Reflection Agent}. The Configuration Agent generates candidate antenna configurations by selecting the electrical downtilt and horizontal and vertical HPBWs of each BS from a validated configuration library, guided by scene information and accumulated optimization experience. The Reflection Agent jointly analyzes ray-tracing evaluation results and scene information to diagnose coverage deficiencies and interference, producing actionable guidance and reusable experience for subsequent configuration updates.

Section \ref{subsectionconfigurationagent} introduces the feasible antenna configuration library and the candidate generation mechanism of the Configuration Agent. Section \ref{subsectionraytracingevaluation} describes the Sionna-based ray-tracing evaluation of candidate configurations, including the calculation of user SINRs and network outage probability. In Section~\ref{subsectionreflectionagent}, the workflow of the Reflection Agent for performance diagnosis and feedback generation are presented. Finally, Section~\ref{subsectionclosedloopoptimizationprocedure} integrates these components into a closed-loop optimization procedure.

\begin{figure*}[t!]
	\centering
	\includegraphics[width=0.95\textwidth]{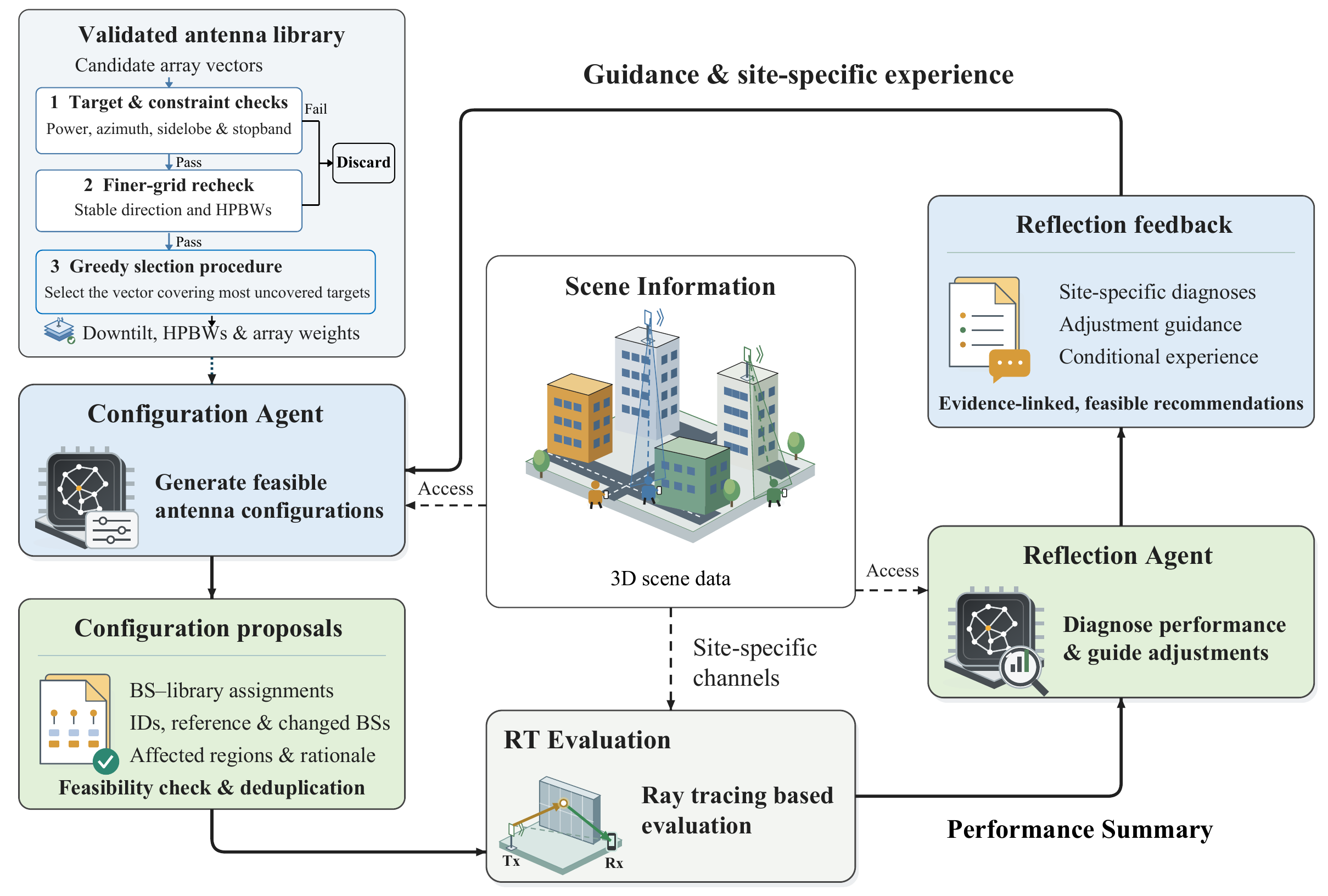}
	\caption{Overall flowchart of the proposed agentic framework}
	\label{overallflowchart}
    \vspace*{-5mm}
\end{figure*}

\subsection{Configuration Agent}
\label{subsectionconfigurationagent}

\subsubsection{Feasible Antenna Configuration Library Construction}

The Configuration Agent requires an action space in which each antenna setting corresponds to an executable array configuration. However, independently specifying the downtilt and horizontal and vertical HPBWs does not guarantee their joint realizability, since the achievable radiation pattern depends on the finite array aperture, element pattern, and excitation constraints. Moreover, configurations with similar mainlobe parameters may exhibit different sidelobe characteristics, leading to different interference levels. Thus, we construct a configuration library that associates each retained parameter triplet with explicit array weights and a numerically validated radiation pattern.

Let $\mathbf{a}=[\theta,\beta^{\mathrm h},\beta^{\mathrm v}]^{\mathsf T}$ denote a target antenna setting. Based on (\ref{eq:realized_antenna_gain}), for the fixed array geometry, the standard array-pattern representation can be expressed as
\begin{equation}
    G(\alpha,\phi;\mathbf{w})
    =
    G_{\mathrm e}(\alpha,\phi)
    \left|
    \sum_{m=1}^{M}
    w_m
    e^{j\frac{2\pi}{\lambda}
    \mathbf{r}_m^{\mathsf T}\mathbf{u}(\alpha,\phi)}
    \right|^2,
    \label{eq:library_array_pattern}
\end{equation}
where $\alpha$ and $\phi$ denote elevation and azimuth relative to the panel, respectively, with
$\mathbf{u}(\alpha,\phi)
=[\cos\alpha\cos\phi,\cos\alpha\sin\phi,\sin\alpha]^{\mathsf T}$.
Independent amplitude and phase control and negligible mutual coupling are assumed.

We first sample target triplets within the prescribed parameter bounds. Initial weight vectors are generated by combining phase steering with uniform or Gaussian amplitude tapers over different horizontal and vertical active apertures. For targets without an acceptable initial realization, these vectors initialize a local pattern-synthesis problem. Specifically, according to (\ref{eq:target_pattern}), the reference power pattern can be expressed as 
\begin{equation}
\begin{aligned}
    T_{\mathbf{a}}(\alpha,\phi)
    &=10^{-A_{\mathbf{a}}(\alpha,\phi)/10},
\end{aligned}
    \label{eq:library_target_pattern}
\end{equation}
where $A_{\mathbf{a}}(\alpha,\phi)
    =\min\left\{
    12\left(\frac{\phi}{\beta^{\mathrm h}}\right)^2
    +12\left(\frac{\alpha+\theta}{\beta^{\mathrm v}}\right)^2,
    A_{\max}
    \right\}$. The excitation weights are refined by solving
\begin{equation}
\begin{aligned}
    \underset{\mathbf{w}}{\operatorname{minimize}}
    \quad&
    \sum_{q=1}^{Q}\rho_q
    \left[
    \frac{G(\alpha_q,\phi_q;\mathbf{w})}{G_0(\mathbf{w})}
    -T_{\mathbf{a}}(\alpha_q,\phi_q)
    \right]^2\\
    \operatorname{subject~to}\quad&
    \|\mathbf{w}\|_2^2=1,\\
    &|w_m|^2\leq p_{\max},\quad m=1,\ldots,M,\\
    &G_0(\mathbf{w})\geq G_{\min},\\
    &G(\alpha_q,\phi_q;\mathbf{w})
    \leq \eta G_0(\mathbf{w}),
    \quad q\in\mathcal{Q}_{\mathrm{out}},
\end{aligned}
    \label{eq:library_pattern_synthesis}
\end{equation}
where $G_0(\mathbf{w})=G(-\theta,0;\mathbf{w})$, and
$\{(\alpha_q,\phi_q)\}_{q=1}^{Q}$ is the synthesis grid.
The nonnegative weights $\rho_q$ emphasize mainlobe accuracy,
$p_{\max}$ limits the power fraction assigned to each element,
and $G_{\min}>0$ prevents a vanishing response in the target direction.
The set $\mathcal{Q}_{\mathrm{out}}$ contains stopband samples separated from the mainlobe by a transition region, and $\eta\in(0,1)$ limits their relative power.
We solve \eqref{eq:library_pattern_synthesis} locally from multiple initializations using sequential quadratic programming.
The normalization by $G_0(\mathbf{w})$ is used only for pattern fitting; the resulting configurations retain their actual gain levels.

Each candidate undergoes independent validation on a finer angular grid. From the dominant front-facing peak $(\hat{\alpha},\hat{\phi})$, we obtain the realized downtilt $\hat{\theta}=-\hat{\alpha}$. The realized HPBWs $\hat{\beta}^{\mathrm h}$ and $\hat{\beta}^{\mathrm v}$ are measured between the nearest half-power crossings on the azimuth and elevation cuts through this peak, respectively. A candidate matches its target only if
\begin{equation}
    \max\left\{
    \frac{|\hat{\theta}-\theta|}{\epsilon_{\theta}},
    \frac{|\hat{\beta}^{\mathrm h}-\beta^{\mathrm h}|}
         {\epsilon_{\mathrm h}},
    \frac{|\hat{\beta}^{\mathrm v}-\beta^{\mathrm v}|}
         {\epsilon_{\mathrm v}}
    \right\}\leq 1,
    \label{eq:library_matching_tolerance}
\end{equation}
where $\epsilon_{\theta}$, $\epsilon_{\mathrm h}$, and
$\epsilon_{\mathrm v}$ are prescribed tolerances.
As shown in Fig.~\ref{overallflowchart}, in addition to matching the target downtilt and HPBWs within the prescribed tolerances, each candidate should also satisfy the parameter bounds, power constraints, azimuth alignment tolerance, and sidelobe and stopband limits. Because the mainlobe direction and half-power crossings are estimated from discrete angular samples, the measurements are re-evaluated on finer grids to ensure that the estimated direction and beamwidths remain stable. Any candidates with anbigous mainlobe or an unresolved beamwidth is discarded.

Multiple validated vectors may realize the same target combination of downtilt and HPBWs, and a single vector may realize multiple targets. To minimize the library size, we employ a greedy selection procedure, iteratively choosing the vector that covers the largest number of remaining unrepresented targets at each step. Once every successfully realized target has a representative, we remove any selected vector whose targets are entirely represented by other selected vectors. The retained vectors and their realized antenna parameters form the library $\mathcal{L}$ as follows,
\begin{equation}
    \mathcal{L}
    =
    \left\{
    \left(\hat{\mathbf{a}}_{\ell},\mathbf{w}_{\ell}\right)
    \right\}_{\ell=1}^{L},
    \qquad
    \hat{\mathbf{a}}_{\ell}
    =
    \begin{bmatrix}
    \hat{\theta}_{\ell}\\
    \hat{\beta}^{\mathrm h}_{\ell}\\
    \hat{\beta}^{\mathrm v}_{\ell}
    \end{bmatrix},
    \label{eq:validated_configuration_library}
\end{equation}
where $L$ is the number of retained configurations.
Each supported target is assigned a representative entry using a fixed rule based on parameter error and pattern quality.
The library provides a finite validated subset of achievable settings under the adopted array model.
Its construction does not use urban-scene channel data or network outage values, allowing reuse across scenes with the same antenna and frequency settings.
During online optimization, the Configuration Agent selects library entries for the BSs, while their site-specific network performance is determined by the subsequent ray-tracing evaluation.

\subsubsection{Antenna Configuration Proposal Generation}

Based on the validated antenna configuration library, the Configuration Agent generates candidate antenna assignments using site-specific evaluation evidence. For a network with $B$ BSs, an assignment is represented by $\mathbf{x}=[\ell_1,\ldots,\ell_B]^{\mathsf T}$, where $\ell_b\in\{1,\ldots,L\}$ identifies the library entry assigned to BS $b$. At iteration $t$, the Configuration Agent receives the fixed site information, the outage objective, the best evaluated assignment $\mathbf{x}_t^\star$, previous evaluation results, and the latest diagnosis from the Reflection Agent. Library entries are accessed through a compact catalog and targeted retrieval tools. The generation procedure comprises diagnostic adjustment, coordinated reconfiguration, and systematic probing.

Specifically, the agent uses regional outage statistics and signal--interference diagnostics to identify candidate BSs for adjustment and formulate the required parameter changes. Starting from $\mathbf{x}_t^\star$, the agent typically modifies one or two BSs while retaining the remaining assignments. Each modification selects a validated library entry according to its realized downtilt and beamwidths. Since these parameters can be coupled in a realizable antenna pattern, the complete parameter change is considered when assessing whether a library transition supports the intended adjustment. 

Furthermore, to investigate interactions among BSs, the Configuration Agent also generates joint assignments by combining settings from an archive of well-performing, evaluated network configurations with less-tested library entries. Archived settings are reused at their corresponding BSs to retain their site-specific context. The agent considers coverage overlap and interference relationships when selecting simultaneous changes.

Moreover, a deterministic scheduler within the Configuration Agent tracks the tested adjustment directions for each BS. It periodically reserves an evaluation slot for an under-tested feasible transition involving downtilt, horizontal beamwidth, vertical beamwidth, or coupled parameter changes. It also tracks pending reflection recommendations.

At iteration $t$, the Configuration Agent returns a batch of structured proposals, which can be represented as 
\begin{equation}
    \mathcal{Q}_t
    =
    \left\{\mathcal{D}_{t,k}\right\}_{k=1}^{K_t},
    \label{eq:antenna_proposal_batch}
\end{equation}
where $K_t$ is the number of proposals admitted under the available evaluation budget. The $k$-th proposal can be represented as
\begin{equation}
    \mathcal{D}_{t,k}
    =
    \big(
        i_{t,k},
        p_{t,k},
        \mathbf{x}_{t,k},
        \mathcal{S}_{t,k},
        \mathcal{R}_{t,k},
        \rho_{t,k}
    \big),
    \label{eq:structured_antenna_proposal}
\end{equation}
where $i_{t,k}$ is a unique proposal identifier and $p_{t,k}$ identifies the evaluated reference from which the proposed changes are defined. $\mathbf{x}_{t,k}$ specifies an executable configuration for every BS, which can be expressed as 
\begin{equation}
    \mathbf{x}_{t,k}
    =
    [\ell_{t,k,1},\ldots,\ell_{t,k,B}]^{\mathsf T},
    \qquad
    \ell_{t,k,b}\in\{1,\ldots,L\}.
    \label{eq:proposal_antenna_assignment}
\end{equation}

Let $\mathbf{x}_{t,k}^{\mathrm{ref}}$ denote the assignment associated with $p_{t,k}$, with its $b$-th entry denoted by $\ell_{t,k,b}^{\mathrm{ref}}$. The modified BS set is
\begin{equation}
    \mathcal{S}_{t,k}
    =
    \left\{
        b\in\{1,\ldots,B\}
        \,\middle|\,
        \ell_{t,k,b}\neq\ell_{t,k,b}^{\mathrm{ref}}
    \right\}.
    \label{eq:proposal_modified_bs_set}
\end{equation}
The set $\mathcal{R}_{t,k}$ contains the identifiers of the service regions expected to be affected by these modifications. The structured rationale $\rho_{t,k}$ records the supporting evaluation evidence, the intended parameter changes, and the anticipated coverage and interference effects, including possible performance degradation in other regions. These anticipated effects are configuration hypotheses to be assessed by the subsequent evaluator.

Before evaluation, a deterministic validator checks assignment completeness, library membership, reference validity, and consistency of the reported modifications. Duplicate assignments are identified using $\mathbf{x}_{t,k}$ and removed against both the evaluation history and the current batch. Invalid or duplicate proposals are returned for correction. The accepted assignments are then passed to the RT evaluator, which will be discussed in the following subsection.

\subsection{Ray Tracing Evaluation}
\label{subsectionraytracingevaluation}

The RT evaluator assesses the performance of the antenna configuration proposals in the target scene using Sionna RT~\cite{aoudia2025sionnart}. For each proposal $\mathcal{D}_{t,k}$, its assignment $\mathbf{x}_{t,k}$ determines the excitation weights retrieved from the validated library. The evaluator combines these weights with the site-specific channel responses obtained from the three-dimensional scene. It then computes the received powers and calculates the SINR and outage probability according to (\ref{eq:downlink_sinr}) and (\ref{eq:outage_probability}). 

The evaluation results can be summarized as a structured performance summary $\mathbf{y}_{t,k}$. Its primary global metric is $J_{t,k}=P_{\mathrm{out}}(\mathbf{x}_{t,k})$, which determines the quality of the proposed configuration. However, this threshold-based metric does not capture SINR changes that leave users on the same side of the outage threshold.
To provide additional diagnostic information, the summary includes the fifth and tenth SINR percentiles, characterizing the lower-tail SINR performance. It also records the number of users within $1$~dB of the outage threshold, identifying users whose coverage status may be sensitive to subsequent antenna adjustments.

To localize the observed performance changes, the summary further includes BS-level and regional statistics, such as associated user counts, regional outage probabilities, and
desired-signal and interference powers. These statistics help identify underserved regions and assess how antenna adjustments affect both useful signal strength and inter-cell interference, providing evidence for site-specific reflection and subsequent configuration generation.

After the structured evaluation procedure, the evaluated proposal batch at iteration $t$ is represented as
\begin{equation}
    \mathcal{Y}_t
    =
    \left\{
        \left(\mathcal{D}_{t,k},\mathbf{y}_{t,k}\right)
        : k=1,\ldots,K_t
    \right\}.
    \label{eq:evaluated_antenna_proposal_batch}
\end{equation}
The proposal with the lowest outage probability in the
current batch is selected as
\begin{equation}
    k_t^\star
    \in
    \arg\min_{k\in\{1,\ldots,K_t\}}
    J_{t,k}.
    \label{eq:best_evaluated_antenna_proposal}
\end{equation}
The proposal and performance summary are denoted by $\mathcal{D}_t^\star=\mathcal{D}_{t,k_t^\star}$ and $\mathbf{y}_t^\star=\mathbf{y}_{t,k_t^\star}$, respectively. Then, the evaluated batch and the selected proposal are passed to the Reflection Agent. Using these numerical results and the scene information, the Reflection Agent assesses the configuration hypotheses, infers likely site-specific causes of the observed performance, and formulates guidance for subsequent antenna adjustments.

\subsection{Reflection Agent}
\label{subsectionreflectionagent}

The Reflection Agent interprets the evaluated configuration proposals and converts their numerical outcomes into site-specific diagnoses, configuration adjustment guidance, and reusable experience. Specifically, its input at iteration $t$ can be represented as
\begin{equation}
    \mathcal{C}_t^{\mathrm{ref}}
    =
    \big(
        \mathcal{Y}_t,
        \mathcal{D}_t^\star,
        \mathbf{y}_t^\star,
        \mathcal{V}_t,
        \mathcal{L}_t,
        \mathcal{A}_t
    \big),
    \label{eq:reflection_agent_context}
\end{equation}
where $\mathcal{Y}_t$ contains the evaluated proposals and their performance summaries, and $(\mathcal{D}_t^\star,\mathbf{y}_t^\star)$ identifies the best proposal in the current batch.
The record collection $\mathcal{V}_t$ provides the evaluated references identified by the proposal identifiers and the current best retained configuration. The library view $\mathcal{L}_t$ contains retrieved entries from the fixed validated library, including their identifiers and realized antenna parameters. In addition, $\mathcal{A}_t$ records the execution status and available outcomes of previous adjustment recommendations, allowing the agent to distinguish completed evaluations from pending suggestions.

Based on this context, the output of the Reflection Agent can be expressed as
\begin{equation}
    \mathcal{R}_t
    =
    \big(
        \mathcal{I}_t,
        \mathcal{G}_t,
        \Delta\mathcal{M}_t
    \big),
    \label{eq:reflection_agent_output}
\end{equation}
where $\mathcal{I}_t$ contains the site-specific diagnoses, $\mathcal{G}_t$ specifies configuration adjustment guidance, and $\Delta\mathcal{M}_t$ contains newly extracted experience.

Specifically, for $\mathcal{I}_t$, the Reflection Agent compares each proposal with its stated reference using the same user samples and evaluation settings. It examines rescued, newly lost, persistently uncovered, and near-threshold users to identify the regions responsible for the observed outage changes. Desired-signal, interference, and association statistics are then used to formulate explanations consistent with these observations. For example, rescued users whose serving BS and desired power remain unchanged, but whose interference decreases, provide evidence of interference relief. Each diagnosis records the affected BSs and regions, the numerical observation, a mechanism hypothesis, and references to the supporting evaluation results.

The guidance $\mathcal{G}_t$ translates the diagnoses into prioritized, feasible tests for the Configuration Agent. Each recommendation specifies its reference evaluation, the BSs to modify, the replacement library entries, the applicable conditions, and the expected performance effect or uncertainty to be resolved. Recommendations may involve a controlled adjustment at one BS, a coordinated change at two BSs, or a test of an under-examined parameter direction. The associated rationale identifies the intended benefit and possible coverage losses in other regions. Their performance is subsequently assessed by the RT evaluator.

The memory increment $\Delta\mathcal{M}_t$ abstracts conditional observations from completed evaluations. Each entry retains the relevant site and configuration conditions, the antenna adjustment, the observed outcome, and its supporting evidence. For example, an observed benefit of reducing horizontal beamwidth is stored together with the corresponding BS, affected regions, and neighboring BS configurations. After checking evidence references and configuration
identifiers, the new experience is added to the persistent
memory, with duplicate entries removed.


\subsection{Closed-Loop Optimization Procedure}
\label{subsectionclosedloopoptimizationprocedure}


In this subsection, we summarize the above three parts and formulate a closed-loop agentic optimization of antenna configurations. Specifically, the proposed agentic framework is summarized in Fig.~\ref{overallflowchart} and Algorithm~\ref{alg:closed_loop_antenna_configuration}.
The Configuration Agent, RT evaluator, and Reflection Agent interact through a shared optimization state, allowing evaluated outcomes to guide subsequent proposals while preserving the best retained configuration.

\begin{algorithm}[t]
\caption{Agentic Optimization of Antenna Configurations}
\label{alg:closed_loop_antenna_configuration}
\small
\begin{algorithmic}[1]
\REQUIRE Scene information, validated library $\mathcal{L}$, initial feasible proposal $\mathcal{D}_0$, maximum iterations $T$
\ENSURE Best retained antenna configuration
\STATE Evaluate $\mathcal{D}_0$ to obtain $\mathbf{y}_0$
\STATE $\mathcal{B}_1\gets(\mathcal{D}_0,\mathbf{y}_0)$,
$\mathcal{H}_1\gets\{\mathcal{B}_1\}$, $\mathcal{M}_1 \gets$ $\varnothing$, $\mathcal{R}_0\gets\varnothing$, $\mathcal{A}_1\gets\varnothing$, $t\gets1$, $n_1\gets1$, $c_1\gets0$
\STATE Set $\Xi_1$ according to (\ref{eq:closed_loop_state})
\WHILE{$t\leq T$}
    \STATE Construct the Configuration Agent context from $\Xi_t$ and $\mathcal{L}$
    \STATE Generate $\mathcal{Q}_t$ according to (\ref{eq:antenna_proposal_batch})
    \STATE Evaluate $\mathcal{Q}_t$ to obtain $\mathcal{Y}_t$ in (\ref{eq:evaluated_antenna_proposal_batch})
    \STATE Select $(\mathcal{D}_t^\star,\mathbf{y}_t^\star)$ according to \eqref{eq:best_evaluated_antenna_proposal}
    \STATE $\mathcal{B}_{t+1}\gets
    \operatorname{Best}(\mathcal{B}_t,\mathcal{Y}_t)$ according to \eqref{eq:closed_loop_state_updates}
    \STATE Resolve previous action outcomes and construct
    $\mathcal{C}_t^{\mathrm{ref}}$ in (\ref{eq:reflection_agent_context}), including
    $\mathcal{B}_{t+1}$
    \STATE Obtain $\mathcal{R}_t$ according to (\ref{eq:reflection_agent_output})
    \STATE Update $\mathcal{H}_{t+1}$, $\mathcal{M}_{t+1}$, and $\mathcal{A}_{t+1}$ according to \eqref{eq:closed_loop_state_updates}
    \STATE $t\gets t+1$
\ENDWHILE
\STATE \textbf{return} the best retained antenna configuration
\end{algorithmic}
\end{algorithm}

The procedure starts from a feasible proposal $\mathcal{D}_0$, whose antenna assignments are selected from the validated library $\mathcal{L}$. At the beginning of iteration $t$, the optimization state can be represented as
\begin{equation}
    \Xi_t=
    \big(
        \mathcal{B}_t,
        \mathcal{H}_t,
        \mathcal{M}_t,
        \mathcal{R}_{t-1},
        \mathcal{A}_t
    \big),
    \label{eq:closed_loop_state}
\end{equation}
where $\mathcal{B}_t$ is the best retained proposal and its evaluation record, $\mathcal{H}_t$ is the evaluation history, $\mathcal{M}_t$ stores conditional site-specific experience, and $\mathcal{R}_{t-1}$ is the output of the Reflection Agent in the $(t-1)$-th iteration. The action record $\mathcal{A}_t$ tracks recommendation execution.

Based on (\ref{eq:antenna_proposal_batch}), the Configuration Agent uses the best retained configuration, evaluation history, reflection guidance, and retrieved experience to construct the next proposal batch. The batch size is limited by the remaining configuration budget. A deterministic validator checks complete BS assignments and library membership, and removes configurations already present in the history or current batch. 
Then, the RT evaluator computes each accepted proposal's performance and its changes relative to the stated reference. These results form $\mathcal{Y}_t$ in (\ref{eq:evaluated_antenna_proposal_batch}) and identify the iteration-best output $\mathcal{D}_t^\star$ and $\mathbf{y}_t^\star$ according to (\ref{eq:best_evaluated_antenna_proposal}). Next, based on input in (\ref{eq:reflection_agent_context}), the Reflection Agent returns diagnoses, prioritized configuration tests, and conditional experience as defined in (\ref{eq:reflection_agent_output}). 

During the iteration, the records $\Xi_{t+1}$ are updated as
\begin{equation}
    \begin{aligned}
        \mathcal{B}_{t+1}
        &=
        \operatorname{Best}
        (\mathcal{B}_t,\mathcal{Y}_t),\\
        \mathcal{H}_{t+1}
        &=
        \operatorname{Append}
        (\mathcal{H}_t,\mathcal{Y}_t),\\
        \mathcal{M}_{t+1}
        &=
        \operatorname{Dedup}
        (\mathcal{M}_t\cup\Delta\mathcal{M}_t),\\
        \mathcal{A}_{t+1}
        &=
        (\mathcal{A}_t,\mathcal{Y}_t,\mathcal{G}_t;
         \mathcal{B}_{t+1}).
    \end{aligned}
    \label{eq:closed_loop_state_updates}
\end{equation}
where $\operatorname{Append}$ preserves all evaluated proposals, the $\operatorname{Dedup}$ operator removes duplicate experience records and $\operatorname{Best}$ retains the best record. The loop terminates when the iteration limit $T$ is reached. 

\textbf{Remark 1} (Extensibility of the proposed framework): The proposed framework follows a modular closed loop comprising candidate generation, feasibility validation, performance evaluation, reflection, and experience updating. This structure allows the framework to be adapted to other site-specific tasks, such as resource allocation and network planning. Adaptation requires task-specific definitions of the action space, constraint checks, evaluation metrics, and diagnostic information.

To clarify the closed-loop procedure of the agentic framework, here we give an illustrative example, where an urban scenario in which users on a street served by BS-1 experience strong interference from BS-2 is considered. Based on the evaluation results, the Reflection Agent identifies BS-2 as a dominant interferer and recommends testing a larger downtilt angle to reduce radiation toward the affected street. The Configuration Agent selects a corresponding entry from the validated library and generates an updated proposal. The RT evaluator then recomputes network performance to assess interference reduction and possible coverage losses. If the outage improvement is confirmed, the updated configuration is retained, and the observed outcome is stored as conditional site-specific experience to guide subsequent adjustments.

\section{Simulation Results}
\label{sectionIV}

In this section, simulation results are presented to verify the performance of our proposed agentic scheme.

\subsection{Simulation Setup}

\begin{figure}[t!]
	\centering
	\includegraphics[width=0.49\textwidth]{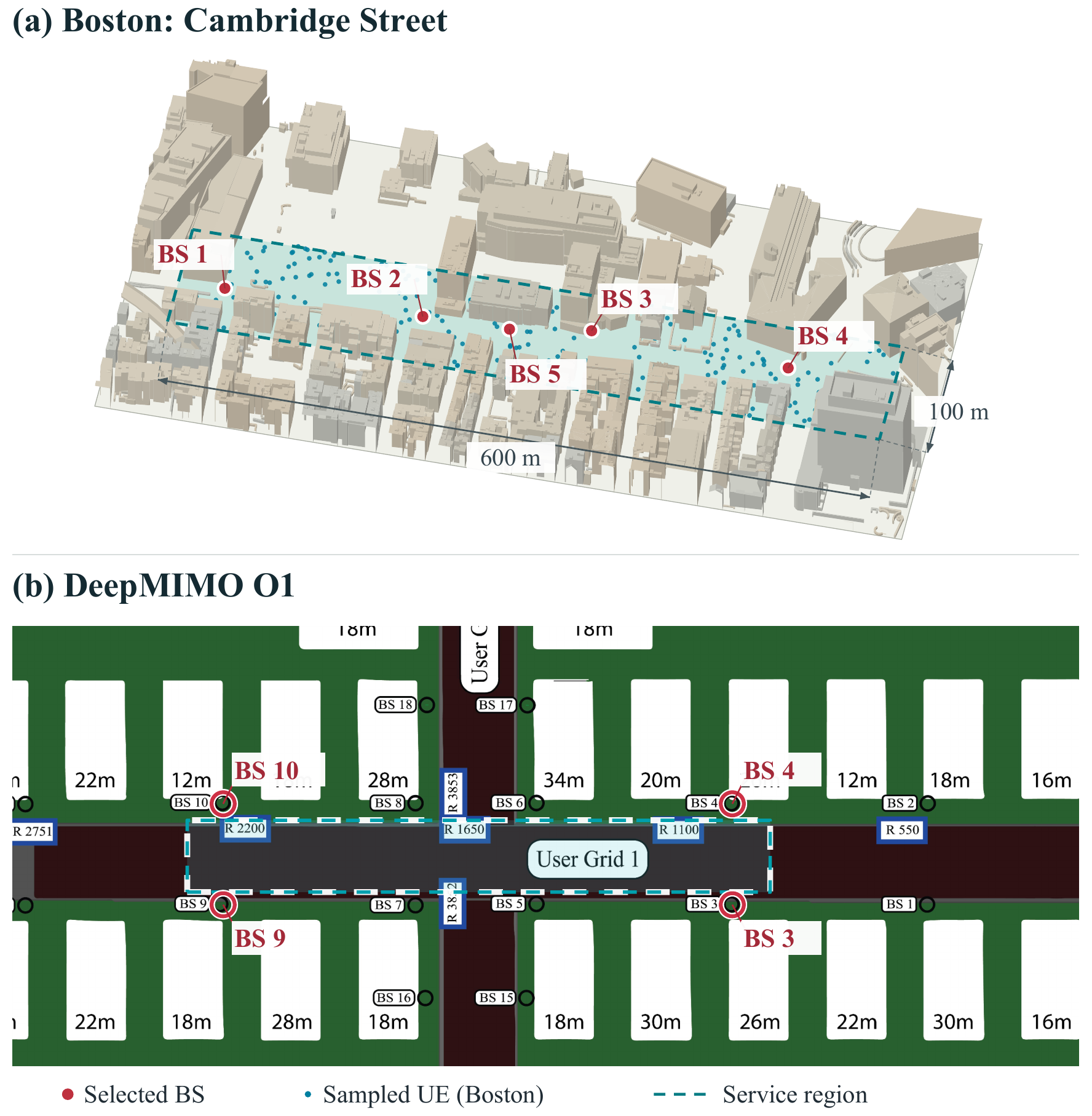}
	\caption{Simulation scenarios: (a) Boston; (b) DeepMIMO O1.}
	\label{fig:simulation_scenarios}
    \vspace{-5mm}
\end{figure}

We consider two outdoor urban scenarios, namely a street corridor in Boston constructed from BostonTwin~\cite{testolina2024bostontwin} and the O1 scenario from DeepMIMO~\cite{alkhateeb2019deepmimo}, as illustrated in Fig.~\ref{fig:simulation_scenarios}. In the Boston scenario, five BSs serve a $600~\mathrm{m}\times100~\mathrm{m}$ region along Cambridge Street, with their horizontal locations obtained from municipal antenna records and their installation heights assumed to be $10$~m above ground. In the O1 scenario, four original BSs, denoted by TX\_3, TX\_4, TX\_9, and TX\_10, serve a $300~\mathrm{m}\times37~\mathrm{m}$ region, retaining the original BS and UE heights of $6$ and $2$~m, respectively. 
For each scenario, we independently sample two sets of $1{,}000$ UE locations within the same service area using the same spatial sampling model. The optimization set is used to optimize the antenna configuration, whereas the evaluation set is used to examine whether the optimized configuration maintains effective coverage under a different realization of UE locations. We evaluate 10 independent evaluation sets and final adopt the arithmetic mean over these 10 sets.

The main simulation parameters and parameters for antenna configuration libraries in Section~\ref{subsectionconfigurationagent} are summarized in Table~\ref{tab:simulation_parameters} and Table~\ref{tab:antenna_library_parameters}, respectively.
Each BS employs an $8\times8$ UPA with half-wavelength element spacing and vertically polarized elements following the 3GPP TR~38.901 antenna model~\cite{3gpp2024tr38901}. All BSs operate over the same $10$~MHz band with a fixed total transmit power of $30$~dBm per BS. Boston channels are generated using Sionna RT~\cite{hoydis2023sionna}, where $3.2\times10^7$ samples are traced per transmitter with a maximum path-interaction depth of 10. O1 channels are reconstructed from the original Wireless InSite propagation paths provided by DeepMIMO. For O1, the path amplitudes are assumed constant over the evaluation bandwidth, while the frequency-dependent phases are reconstructed from the path delays. Received link powers are averaged over $64$ uniformly spaced frequency samples, and each UE associates with the BS providing the largest average received power. An outage occurs when the resulting SINR falls below $0$~dB. Both the Configuration Agent and the Reflection Agent are implemented through GPT-5.6-Sol and Codex SDK with distinct role instructions. Each request combines role-specific instructions, feasibility constraints, a structured description of the current search state, and a context-dependent JSON output schema.

\begin{table}[t]
    \centering
    \caption{Simulation parameters.}
    \label{tab:simulation_parameters}
    \footnotesize
    \renewcommand{\arraystretch}{1.12}
    \setlength{\tabcolsep}{3pt}
    \begin{tabular*}{\columnwidth}{@{\extracolsep{\fill}}lcc@{}}
        \toprule
        \textbf{Parameter} & \textbf{Boston} & \textbf{DeepMIMO O1} \\
        \midrule
        Number of BSs & $5$ & $4$ \\
        Service region (m $\times$ m) & $600\times100$ & $300\times37$ \\
        Carrier frequency & $2$~GHz & $28$~GHz \\
        Evaluation bandwidth & \multicolumn{2}{c}{$10$~MHz} \\
        BS antenna height & $10$~m & $6$~m \\
        UE antenna height & $1.5$~m & $2$~m \\
        Total transmit power per BS & \multicolumn{2}{c}{$30$~dBm} \\
        BS antenna array & \multicolumn{2}{c}{$8\times8$ uniform planar array} \\
        Maximum iterations $T$ & \multicolumn{2}{c}{20} \\
        Element radiation pattern & \multicolumn{2}{c}{3GPP TR~38.901} \\
        Element maximum gain & \multicolumn{2}{c}{$8$~dBi} \\
        Element HPBW (horizontal/vertical)
            & \multicolumn{2}{c}{$65^\circ/65^\circ$} \\
        Antenna polarization & \multicolumn{2}{c}{Vertical} \\
        UE antenna & \multicolumn{2}{c}{Single isotropic element} \\
        SINR outage threshold & \multicolumn{2}{c}{$0$~dB} \\
        Number of frequency samples & \multicolumn{2}{c}{$64$} \\
        Samples per transmitter & \multicolumn{2}{c}{$3.2 \times 10^7$} \\
        Maximum path-interaction depth & \multicolumn{2}{c}{$10$} \\
        Optimization/evaluation UE locations
            & \multicolumn{2}{c}{$1{,}000/1{,}000$} \\
        Receiver noise power & \multicolumn{2}{c}{$-100$~dBm} \\
        \bottomrule
    \end{tabular*}
\end{table}

\begin{table}[t]
    \centering
    \caption{Antenna configuration library parameters.}
    \label{tab:antenna_library_parameters}
    \footnotesize
    \renewcommand{\arraystretch}{1.12}
    \setlength{\tabcolsep}{3pt}
    \begin{tabular*}{\columnwidth}
        {@{\extracolsep{\fill}}lc@{}}
        \toprule
        \textbf{Parameter} & \textbf{Value} \\
        \midrule
        Ranges $(\theta,\beta^{\mathrm h},\beta^{\mathrm v})$
            & $[0^\circ,25^\circ],\ [5^\circ,100^\circ],\
               [5^\circ,65^\circ]$ \\
        Tolerances $(\epsilon_\theta,\epsilon_{\mathrm h},
                     \epsilon_{\mathrm v})$
            & $0.5^\circ,\ \epsilon(\beta^{\mathrm h}),\
               \epsilon(\beta^{\mathrm v})$ \\
        Target attenuation cap $A_{\max}$
            & $30$~dB \\
        Maximum element power fraction $p_{\max}$
            & $0.25$ \\
        Minimum target-direction gain $G_{\min}$
            & $1$ ($0$~dBi) \\
        Maximum stopband power ratio $\eta$
            & $0.1$ ($-10$~dB) \\
        Maximum relative sidelobe level
            & $-10$~dB \\
        Library size $L$
            & $1{,}693$ \\
        \bottomrule
    \end{tabular*}
    \par\smallskip
    \parbox{\columnwidth}{%
        \footnotesize
        The HPBW tolerance is defined as
        $\epsilon(\beta)=\max(0.5^\circ,0.05\beta)$.
    }
\end{table}

Benchmark comparison schemes are considered as follows: (1) Fixed Configuration~\cite{3gpp3dmodelrelease12}, which assigns all BSs the same configuration according to 3GPP default settings, with a downtilt angle, horizontal HPBW, and vertical HPBW of approximately $12^\circ$, $60^\circ$, and $15^\circ$, respectively. (2) Random Search, which randomly samples library entries for individual BSs while excluding previously evaluated network configurations. (3) Coordinate Search, which sequentially optimizes one base station at a time by searching local feasible library neighbors for that BS while holding the configurations of all other BSs fixed, and excluding previously evaluated full-network configurations. (4) BO-based Scheme~\cite{tekgul2024joint}, which employs a Gaussian-process surrogate and expected improvement to select configurations from a sampled candidate pool. All methods share the same BS locations, feasible antenna library, user populations, cached propagation channels, and outage definition. These schemes start from the same initial configuration and common initial probes, with identical limits on the numbers of RT evaluations and iterations.

\subsection{Performance Analysis}
\label{subsec:performance_analysis}

\begin{figure}[t!]
	\centering
	\includegraphics[width=0.49\textwidth]{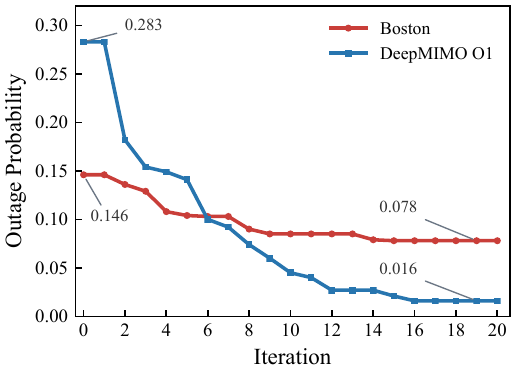}
	\caption{Outage probability versus iteration index in the two scenarios.}
	\label{fig:outage_iterations}
\end{figure}

Fig.~\ref{fig:outage_iterations} illustrates the outage probability of our proposed scheme against the iteration number in two scenarios, with iteration zero representing the initial configuration. In Boston, the outage decreases from 0.146 to 0.108 after four iterations and to 0.090 after eight iterations, reaching 0.078 at iteration 15. In O1, it decreases from 0.283 to 0.149 after four iterations and to 0.074 after eight iterations, reaching 0.016 at iteration 16. These results correspond to relative reductions of $46.6\%$ and $94.3\%$, or net decreases of 68 and 267 users in outage among the 1,000 optimization users, respectively. The trajectories show that successive configuration proposals can improve coverage even after several iterations without progress. Their monotonicity follows from retaining the best confirmed configuration

\begin{figure}[t!]
	\centering
	\includegraphics[width=0.49\textwidth]{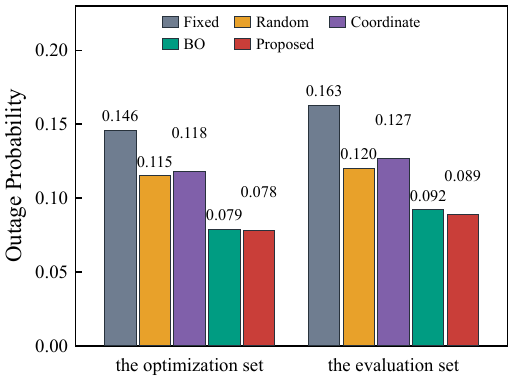}
	\caption{Performance comparison between the proposed scheme and the benchmarks in the Boston scenario}
	\label{fig5}
\end{figure}

\begin{figure}[t!]
	\centering
	\includegraphics[width=0.49\textwidth]{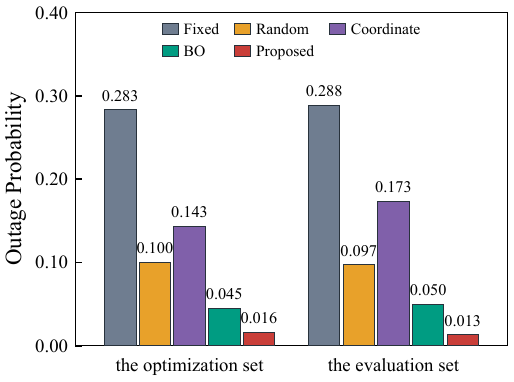}
	\caption{Performance comparison between the proposed scheme and the benchmarks in the DeepMIMO O1 scenario}
	\label{fig6}
\end{figure}

Moreover, the outage probabilities performance of different schemes is shown in Fig.~\ref{fig5} and Fig.~\ref{fig6}, respectively. Specifically, in the Boston scenario, the proposed framework achieves an outage probability of 0.078 on the optimization set, compared with 0.146, 0.115, and 0.118 for fixed configuration, random search, and coordinate search, respectively. The corresponding relative reductions are $46.6\%$, $32.2\%$, and $33.9\%$. The improvements over random and coordinate search indicate that the proposed interaction between configuration generation and performance feedback provides useful guidance for selecting antenna settings in the evaluated scene. In addition, the BO scheme achieves $0.079$, almost the same performance with the proposed agentic scheme in this comparison. Beyond the final objective value, the proposed framework records spatial diagnoses and site-specific configuration experience linked to evaluation evidence, which may help network engineers interpret coverage-interference tradeoffs and guide subsequent tuning. The scalar-objective BO baseline does not explicitly provide such diagnostic guidance, which would require an additional analysis component. Besides, for the evaluation set, although the proposed configuration’s outage increases by 1.1 percentage points relative to the optimization set, it retains its advantage over the considered benchmarks despite the change in user locations. This provides evidence of spatial generalization across sampled user positions within the scene. 

In addition, in the DeepMIMO O1 scenario, the proposed framework achieves an outage of 0.016 and 0.013 on the optimization and evaluation sets, respectively. It reduces outage relative to BO by $64.4\%$ on the optimization set and $74.0\%$ on the evaluation set. The coverage improvement is thus maintained under a different realization of UE locations within the O1 service area. Unlike the small gap observed in Boston, the reported O1 experiment exhibits a clear numerical advantage over all four benchmarks.

\begin{figure*}[t!]
	\centering
	\includegraphics[width=0.92\textwidth]{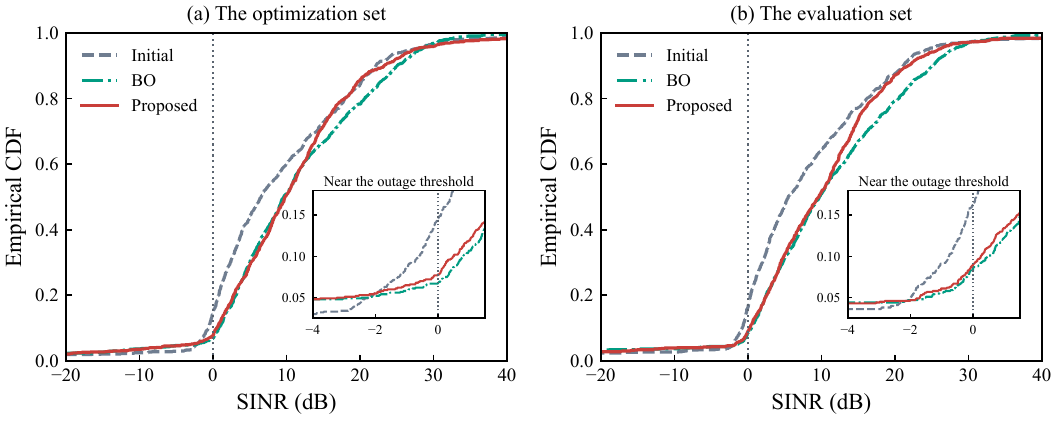}
	\caption{Empirical CDFs of UE SINRs in the Boston scenario, where the insets magnify the SINR distributions near the $0$-dB outage threshold.}
	\label{fig7}
        \vspace*{-4mm}
\end{figure*}

\begin{figure*}[t!]
	\centering
	\includegraphics[width=0.92\textwidth]{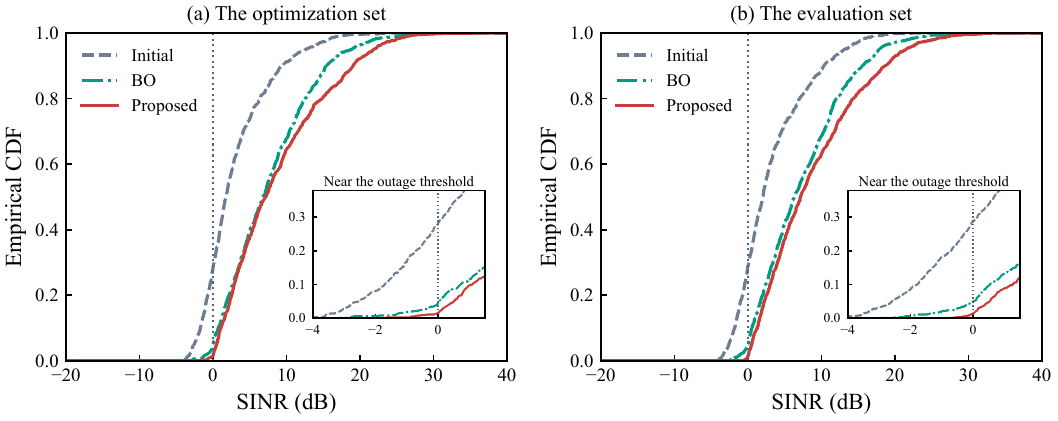}
	\caption{Empirical CDFs of UE SINRs in the DeepMIMO O1 scenario, where the insets magnify the SINR distributions near the $0$-dB outage threshold.}
	\label{fig8}
        \vspace*{-4mm}
\end{figure*}

Fig.~\ref{fig7} and Fig.~\ref{fig8} examine whether the outage reduction is accompanied by broader changes in the user SINR distribution. Specifically, on the Boston's optimization set, the proposed framework increases the median SINR from 6.56 to 9.82 dB and the tenth percentile from -0.60 to 0.51 dB. On the evaluation set, the corresponding values increase from 5.50 to 9.63 dB and from -0.76 to 0.26 dB, respectively. However, the optimization-set fifth percentile decreases from -2.22 to -2.99 dB. Thus, the improvement in aggregate coverage and central SINR statistics coexists with degradation in part of the lower tail. This behavior is consistent with the threshold-based outage objective, which rewards the number of users meeting the SINR requirement without guaranteeing an SINR increase for every user. Besides, in the DeepMIMO O1 scenario, the proposed framework increases the median SINR from 1.78 to 7.20 dB and the tenth percentile from -1.71 to 1.10 dB on the optimization set. On the evaluation set, these statistics improve from 2.05 to 7.16 dB and from -1.78 to 1.28 dB, respectively. Overall, these results indicate that the outage reduction is accompanied by improvements in both lower-tail and median SINR statistics, without implying uniform improvement across all users.

\begin{figure}[t!]
	\centering
	\includegraphics[width=0.49\textwidth]{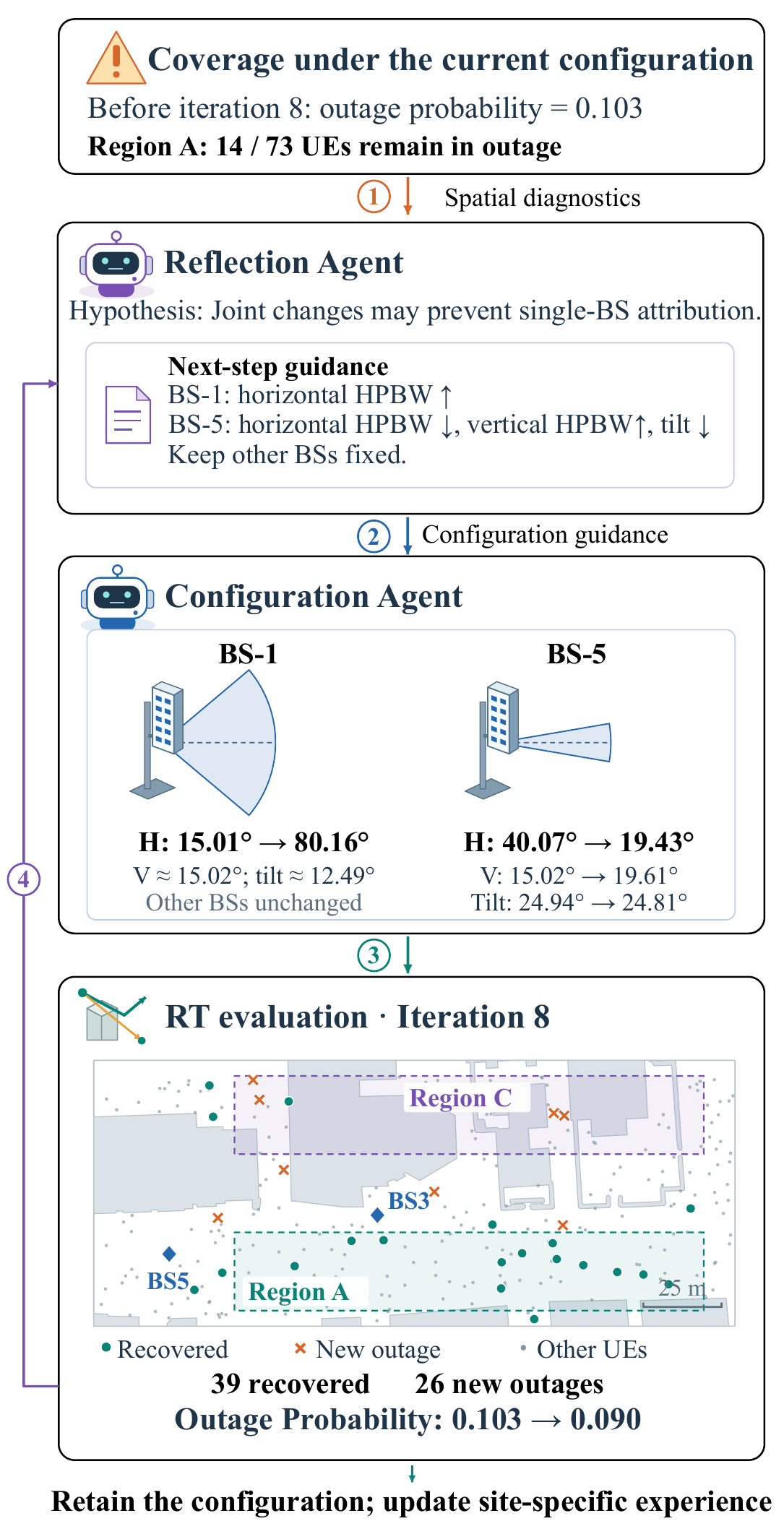}
	\caption{A recorded closed-loop adjustment example at iteration 8 in the Boston scenario.}
	\label{fig9}
\end{figure}

Furthermore, an representative closed-loop adjustment example is illustrated in Fig.~\ref{fig9} to localize the coverage changes.
Starting from an outage probability of $0.103$, the Reflection Agent identifies that 14 of 73 users remain below the SINR threshold and recommends joint adjustments of BS-1 and BS-5. Then, the Configuration Agent selects validated BS-1 and BS-5 entries, including a horizontal HPBW increase from $15.01^\circ$ to $80.16^\circ$ at BS-1 and a decrease from $40.07^\circ$ to $19.43^\circ$ at BS-5. Next, RT evaluation confirms 39 recovered users and 26 new outages, reducing the outage probability to $0.090$, a relative improvement of $12.6\%$. In region A, the number of users in outage decreases from 14 to 2, whereas it increases from 11 to 14 in region C, revealing local regressions despite the overall gain. The configuration is retained, and the measured gains and losses are returned to the Reflection Agent as conditional site-specific experience. This example shows how spatially resolved evaluation informs a testable subsequent adjustment.

\begin{figure}[t!]
	\centering
	\includegraphics[width=0.485\textwidth]{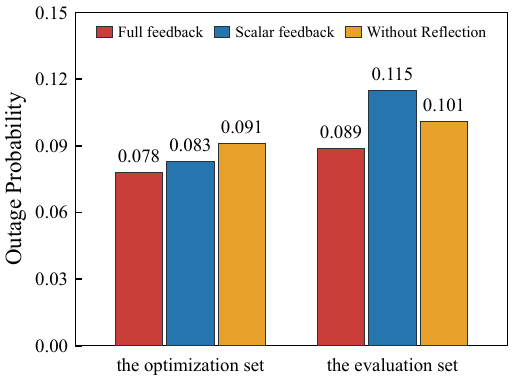}
	\caption{Outage probabilities of the proposed complete framework and two ablation variants in the Boston scenario.}
	\label{fig10}
\end{figure}

\begin{figure}[t!]
	\centering
	\includegraphics[width=0.485\textwidth]{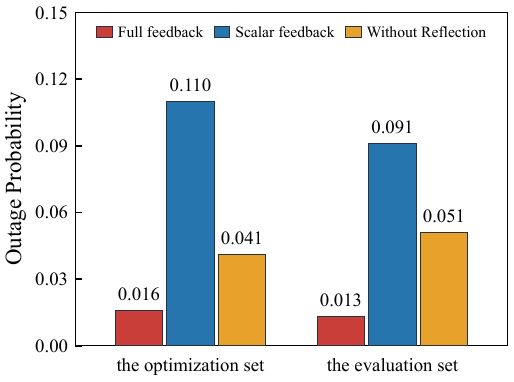}
	\caption{Outage probabilities of the proposed complete framework and two ablation variants in the DeepMIMO O1 scenario.}
	\label{fig11}
\end{figure}

In addition, Fig.~\ref{fig10} and Fig.~\ref{fig11} compare Full Feedback, Scalar Feedback, and the framework without Reflection under the same model and reference-evaluation budget. Full Feedback denotes the complete framework. Scalar Feedback retains global performance summaries and configuration history while omitting spatial diagnostics. Without Reflection removes the Reflection Agent. Specifically, in the Boston scenario, Fig.~\ref{fig10} shows that Full Feedback achieves outage probabilities of 0.078 and 0.089 on the optimization and evaluation sets, respectively. Scalar Feedback yields 0.083 and 0.115, while Without Reflection yields 0.091 and 0.101. Relative to Scalar Feedback, the complete framework reduces outage by $6.0\%$ and $22.6\%$. Relative to Without Reflection, the reductions are $14.3\%$ and $11.9\%$. The larger advantage over Scalar Feedback on the evaluation set is consistent with spatial diagnostics providing useful guidance for configurations that remain effective at independently sampled UE locations. However, Scalar Feedback outperforms Without Reflection only on the optimization set, showing that an improvement in the optimization objective does not necessarily translate into a better evaluation result.

For DeepMIMO O1, Fig.~\ref{fig11} shows a larger separation between the feedback variants. Full Feedback achieves outage probabilities of 0.016 and 0.013, compared with 0.110 and 0.091 for Scalar Feedback and 0.041 and 0.051 for Without Reflection on the optimization and evaluation sets, respectively. These differences correspond to relative reductions of $85.5\%$ and $85.7\%$ over Scalar Feedback, and $61.0\%$ and $74.5\%$ over Without Reflection. Scalar Feedback performs worse than Without Reflection on both sets, indicating that adding a Reflection Agent with aggregate feedback alone does not ensure better configuration proposals. The results support the usefulness of spatially resolved feedback in the complete framework for this experiment, while also showing that the value of reflection depends on the information available to it.

\vspace{-2.5mm}
\section{Conclusions}
\label{sectionV}
This paper has proposed an agentic framework for site-specific cell shaping that optimizes realizable BS antenna configurations to minimize downlink outage probability. Through a closed loop, the Configuration Agent generates proposals from a pattern-validated library, ray-tracing evaluation quantifies their performance, and the Reflection Agent translates evaluation feedback into configuration guidance and accumulated site-specific experience. Simulations in real-world urban scenarios demonstrated lower outage probabilities than the considered benchmarks, highlighting the effectiveness of integrating scene information, agentic reasoning, and iterative reflection for antenna configuration. 
Future work will investigate cooperation among multiple agents for cell shaping~\cite{liumalam4com2026}.

\footnotesize
\bibliographystyle{IEEEtran}
\bibliography{IEEEabrv,reference}

\vfill

\end{document}